\documentclass{elsart}

\usepackage{amssymb}
\usepackage{graphicx}
\usepackage{subfigure}

\def\mystrut{\vrule height 3.5ex depth 2.2ex width 0pt}

\begin{document}

\begin{frontmatter}
\title{Testing confining potentials \\
through meson/baryon hyperfine splittings}

\author{Boaz Keren-Zur}
\address{School of Physics and Astronomy \\
Raymond and Beverly Sackler Faculty of Exact Sciences \\
Tel Aviv University, Tel Aviv, Israel}
\thanks[]{Supported in part by the Israel Science Foundation}
\ead{kerenzu@post.tau.ac.il}

\begin{abstract}
In this paper we discuss a quantity that can be derived from the hadronic
spectrum -- the ratio between the color hyperfine splitting of a $K$ meson and
that of a $\Sigma$ baryon.  It is shown that within the constituent quark
model this ratio depends only on the ratio of contact probabilities 
in the hadrons.
We compute this ratio assuming several phenomenological potential models, and
show that the best agreement with data is obtained from the Cornell potential -
Coulomb + linear. Thus the analysis of color hyperfine interactions corroborates
the preference for the Cornell potential based on quarkonium spectra.
\end{abstract}

\begin{keyword}
QCD \sep Quark potential models \sep Color hyperfine interaction
\end{keyword}

\end{frontmatter}

\section{Introduction}
The recent measurement of the $\Sigma_b$ baryon masses \cite{CDF06}
provides us with a rare opportunity to better our understanding of the
interaction between a heavy quark and a light diquark at low energies. In this
paper we study the dynamics of such a baryon by comparing its properties with
those of a meson which has similar quark content. We focus on their color
hyperfine splitting (HF), which is defined as the mass difference between 2 similar hadrons that differ by one unit of spin (e.g. ${\Delta_K}=M_{K^*}-M_{K}$).
According to the available experimental data, the ratio between the HF splitting of a $\bar{u}q$ meson and that of a $udq$ baryon (where $q$ stands for $s$, $c$ or $b$) is almost independent of the heavy quark flavor:
\begin{equation}
{\Delta_K}/{\Delta_{\Sigma}}\approx{\Delta_D}/{\Delta_{\Sigma_c}}\approx{\Delta_B}/{\Delta_{\Sigma_b}}\approx 2.1
\end{equation}
This quantity was first discussed in \cite{Lipkin86}, but so far we have not been able to 
explain this result from first principles. However, when using phenomenological 
models this ratio takes on a new meaning -- it will be shown below that within
the constituent quark model, 
this ratio is equal to the ratio of contact probabilities between quarks inside the hadron. For example:
\begin{equation}
{\Delta_K}/{\Delta_{\Sigma}}=
\frac{4}{3}\frac{\langle\psi|\delta(\vec{r}_u-\vec{r}_{\bar{s}})|\psi\rangle_K}{\langle\psi|\delta(\vec{r}_u-\vec{r}_s)|\psi\rangle_{\Sigma}}
\label{formula_DK_DSigma}
\end{equation}

The interesting feature of this quantity is that it is determined by two
parameters only: the shape of the potential that binds the quarks together,
and the quark mass ratio. While the constituent quark masses 
can be extracted with sufficient accuracy from relations between other hadron
masses \cite{KarlinerLipkin03}, 
the shape of the potential is still under continuing research.
This subject was studied in the past mainly through the analysis of the 
quarkonium spectrum and the results favored a Coulomb + linear potential
\cite{Cornell78, Cornell80, QuiggRosner79},
but this result still awaits a solid theoretical justification.

We therefore suggest to use this ratio as a new means to test the possible
quark confinement models.  We computed the ratio of contact probabilities
assuming various phenomenological models: harmonic oscillator, 
Coulomb interaction, linear potential, logarithmic potential and the Cornell potential (linear+Coulomb), 
and compared the different results with the available data for the HF splitting ratio.

\section{The constituent quark model}
Low-energy constituent-quark models have been surprisingly successful in predicting hadron masses \cite{KarlinerLipkin03,KarlinerLipkin06}. In such models the hadrons are considered as bound states of constituent quarks in a confining potential, and their total mass is given by the Sakharov-Zeldovich formula
\cite{SakharovZeldovich66,DerujulaDeorgiGlashow75,GasiorowiczRosner81}:
\begin{equation}
M_{hadron}=\sum_i m_i + \sum_{i< j}V^{HF}_{ij}~~,
\end {equation}
where $m_i$ is the constituent quark mass and $V^{HF}_{ij}$ is the color hyperfine (HF) interaction between the quarks $i$ and $j$. The HF interaction is described as the product of the magnetic moments of the quarks, in analogy with the electromagnetic hyperfine interaction, and is given by the expression:
\begin{equation}
V^{HF}_{ij}=v_0\sum_{i > j}\vec{\lambda_i}\cdot\vec{\lambda_j}
\frac{\vec{\sigma_i}\cdot\vec{\sigma_j}}{m_im_j}\langle\delta(r_{ij})\rangle
\label{formula_HF_basic}
\end{equation}
where $\vec{\lambda_i}$ are the $SU(3)$ generators,
$m_i$ and $\sigma_i$ are the mass and spin of the $i$'th quark, $r_{ij}$ is
the distance between the quarks $i$ and $j$, and $v_0$ is the coupling constant
(e.g., in the case of one-gluon exchange between quarks and anti-quarks we have  
$\displaystyle{v_0\vec{\lambda_q}\cdot\vec{\lambda_{\bar{q}}}=\frac{8\pi\alpha_s}{9}}$).

We note that this is a contact interaction and that the formula is correct for S-wave hadrons only. 
We also assume that the same formula and parameterization apply for both mesons and baryons.

\subsection{Extracting constituent quark masses}
We begin by demonstrating calculations with which constituent quark masses can
be extracted from the hadron spectrum. The first example is the extraction of
quark mass differences from $\Lambda$ baryon masses:
\begin{equation}
\Lambda_c-\Lambda=m_c-m_s~~,
\end{equation}
where we used the fact that the HF interaction between the heavy quark and the
light quarks must be zero, due to the anti-symmetry of the $\Lambda$ baryons
under exchange of the light quarks.

We can also use the expression for the meson HF splittings
\footnote{We use here the normalization under which the spin products of a
vector meson and a pseudoscalar are
$(\vec{\sigma_u}\cdot\vec{\sigma_{\bar{s}}})_{s=1}=1 $ and $(\vec{\sigma_u}\cdot\vec{\sigma_{\bar{s}}})_{s=0}=-3$}
\begin{eqnarray}
\label{formula_K_HF}
M_{K^*}-M_K &=&4v_0\frac{\vec{\lambda_u}\cdot\vec{\lambda_{\bar{s}}}}{m_um_s}\langle\delta(r_{u\bar{s}})\rangle
\end{eqnarray}
to extract the ratio between the $s$-quark and $c$-quark masses:
\begin{eqnarray}
\frac{M_{D^*}-M_D}{M_{K^*}-M_K} = \frac{\displaystyle{4v_0\frac{\vec{\lambda_u}\cdot\vec{\lambda_{\bar{c}}}}{m_um_c}\langle\delta(r_{u\bar{c}})\rangle}}
     {\displaystyle{4v_0\frac{\vec{\lambda_u}\cdot\vec{\lambda_{\bar{s}}}}{m_um_s}\langle\delta(r_{u\bar{s}})\rangle}}
\approx \frac{m_s}{m_c}~~,
\label{formula_mass_ratio}
\end{eqnarray}
where we implicitly assumed that the interaction strength $v_0$ is equal in
both cases and that the contact term does not vary greatly. 

Given the simple form of this model, and the fact that we simply ignored the
dependence on the contact probabilities, the consistency of the predictions
given by this approach is surprising. The following values for the
constituent quark masses can be extracted from these calculations \cite{KarlinerLipkin03}:
\begin{eqnarray}
m_u=m_d=& 360& \textup{MeV} \nonumber \\
m_s=    & 540& \textup{MeV} \nonumber \\
m_c=    &1710& \textup{MeV} \nonumber \\
m_b=    &5050& \textup{MeV} 
\label{formula_quark_masses}
\end{eqnarray}

\subsection{The ratio between HF splitting of baryons and mesons}
We now turn to quantities that involve both mesons and baryons, where the contact probabilities are no longer expected to be similar. We choose to analyze the ratio between two HF splitting values
$\displaystyle{(M_{K^*}-M_K)/(M_{\Sigma^*}-M_{\Sigma})}$,
because as will be seen below, some cancellations occur that simplify the final expression.

The $K$ Meson HF splitting is given by (\ref{formula_K_HF}).  The calculation
of the HF splitting of the $\Sigma$ baryon is easy because it gets no
contribution from the $ud$ diquark.  The explanation for that is as follows:
The total spin of the $\Sigma^*$ is $\displaystyle{\frac{3}{2}}$, so all the quark pairs must be at 
relative spin 1. The $\Sigma$ has isospin-1, therefore the 2 light quarks must also have
relative spin 1, and the contribution of the $ud$ diquark to the HF splitting is zero. 

We are left only with the interaction between the light quarks and the heavy one. 
After some spin products calculations we get the following expression for the 
$\Sigma$ HF splitting:
\begin{equation}
\Delta_{\Sigma} = \frac{6v_0}{m_um_s}
\Big[\vec{\lambda_u}\cdot\vec{\lambda_s}\langle\delta(r_{us})\rangle\Big]_{baryon}
\label{formula_sigma_HF}
\end{equation}

Combining equations (\ref{formula_K_HF}), (\ref{formula_sigma_HF}) and (\ref{formula_SU3_generators}), and the relation:
\begin{equation}
[\vec{\lambda_i}\cdot{\vec{\lambda_j}}]_{meson}=2[\vec{\lambda_i}\cdot{\vec{\lambda_j}}]_{baryon}
\label{formula_SU3_generators}
\end{equation}
we reach the result mentioned above:
\begin{equation}
{\Delta_K}/{\Delta_\Sigma}=
\frac{4}{3}\frac{\langle\psi|\delta(r_{u\bar{s}})|\psi\rangle_{meson}}
{\langle\psi|\delta(r_{us})|\psi\rangle_{baryon}}
\label{formula_K_sigma_HF}
\end{equation}
The advantage of this expression over relations such as (\ref{formula_mass_ratio}) (besides the fact that we don't assume anything about the wave function), is that here we have the same quark content in all the interactions involved, and it seems safer to assume that interaction parameters (confinement strength, HF interaction coupling constant) are identical and therefore can cancel out in the ratio. 

Similar expression are obtained when the $s$ quark is replaced with $c$ or $b$ quarks. The available experimental data is given in table \ref{tab_data}: 
\begin{table} [!htbp]
\centering
\begin{tabular}{cccc} \hline 
  Meson & Baryon & $m_3 / m_1$ & HF splitting \\
        &        &             & ratio        \\ \hline \hline 
\mystrut $K$&$\Sigma  $ &$1.33$& $2.08\pm0.01$\\  
\mystrut $D$&$\Sigma_c$ &$4.75$& $2.18\pm0.08$\\ 
\mystrut $B$&$\Sigma_b$ &$14$  & $2.15\pm0.20$\\ \hline 
\end{tabular}
\caption{\small{Experimental data of meson / baryon HF splitting ratios.}}
\label{tab_data}
\end{table}

An interesting comment is that if we simply assumed that the contact probabilities are inversely proportional to the number of quarks in the hadron, then the expected HF splitting ratio would be 2, which is not very far from the measured data.

\section{Confining potentials}
Following is a list of phenomenological potentials that are often used in the literature as models for the confining potential:
\begin{itemize}
\item Harmonic oscillator (HO) -- For sufficiently small deviations from equilibrium, almost every potential looks like a harmonic potential. Of course, as the deviation from equilibrium gets larger, the potential is less reliable. Nevertheless, it is useful as a proof of concept and for establishing orders of magnitude. An additional advantage is that expectation values for 3-body problems can be calculated analytically.
\item Coulomb interaction -- The leading order QCD diagram involves a single gluon exchange, which gives a Coulomb-like potential.
\item Linear potential -- The asymptotic behavior of the QCD confining potential at large distances is linear.
\item Coulomb + Linear (Cornell potential) -- The Coulomb + linear potential is referred to as the Cornell potential, named after the team which studied it in the late 70's \cite{Cornell78,Cornell80}. This potential is a sum of the two asymptotic limits, and is supported by theoretical calculations using Wilson loops \cite{Brambilla94} and renormalon cancellation \cite{sumino03} techniques.
\item Logarithmic potential -- The logarithm function bears a resemblance to a Coulomb+linear function, and is also used as a phenomenological confining potential. Several aspects of the hadronic spectrum are consistent with a logarithmic potential model.
\end{itemize}

The problem of choosing the most effective confining potential model was
studied in the past mainly through the analysis of quarkonium spectra and
leptonic decays \cite{QuiggRosner79}, and the results favored the Cornell
potential. We addressed this issue by computing the contact probability ratio
from Eq.\ (\ref{formula_K_sigma_HF}) given each of the above models, and
comparing with the available measurements.

The quark mass ratios used in the following calculations were taken from Eq. (\ref{formula_quark_masses}). 
Due to the error bars in the experimental data, the level of accuracy required from the computation of contact probabilities is in the order of $1\%$.

\subsection{Harmonic oscillator}
In this model the potential that binds the pairs of quarks takes the form
\begin{equation}
V=\sum_{i< j}\frac{V_{ho}}{2}(\vec{\lambda_i}\cdot\vec{\lambda_j})(\vec{r_i}-\vec{r_j})^2
\end{equation}
where $V_{ho}$ is an effective coupling constant, and the product of the $SU(3)$ generators 
depends on the specific quark configuration in question (meson or baryon).

\subsubsection{HF splitting in a harmonic oscillator model -- mesons}
The ground-state wave function of a harmonic oscillator for a system of two masses $m_1$, $m_2$ is 
\begin{equation}
\psi_0(\vec{r})=(\frac{\alpha_r}{\sqrt{\pi}})^\frac{3}{2}e^{-\frac{1}{2}\alpha_r^2r^2}
\end{equation}
where we use the definitions
\begin{eqnarray}
m_r&\equiv&\frac{m_1m_2}{m_1+m_2} \qquad
\omega_r^2\equiv\frac{V_{ho}}{m_r}(\vec{\lambda_1}\cdot\vec{\lambda_2}) \\ \nonumber 
\alpha_r&\equiv& \sqrt{\frac{m_r\omega_r}{\hbar}}
\end{eqnarray}
leading to
\begin{equation}
\langle \psi_0 | \delta(r_{u\bar{s}}) | \psi_0 \rangle=|\psi_0(0)|^2=(\frac{\alpha_r}{\sqrt{\pi}})^3
\end{equation}

\subsubsection{HF splitting in a harmonic oscillator model -- baryons}
\begin{figure}[!htbp]
	\centering
		{\includegraphics*[width=0.6\textwidth]{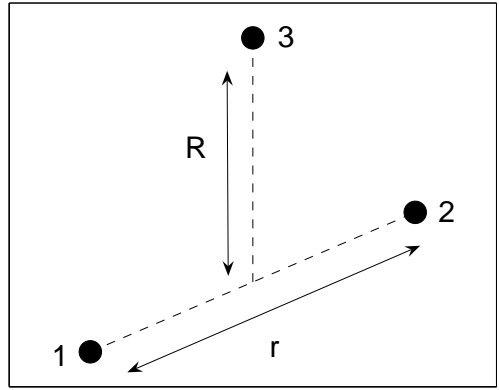}}
	\caption{\small{The two orthogonal modes of the 2+1 harmonic oscillator system}}
	\label{fig:HOmodes}
\end{figure}

The three body HO Hamiltonian, with $m_1=m_2$ and $(\vec{\lambda_1}\cdot\vec{\lambda_3})=(\vec{\lambda_2}\cdot\vec{\lambda_3})$,
can be decoupled by a simple transformation of coordinates:
\begin{equation}
\vec{r}\equiv\vec{r_{1}}-\vec{r_{2}} \qquad \vec{R} \equiv
\vec{r_{3}}-\frac{1}{2}\vec{(r_{1}}+\vec{r_{2}})~~.
\end{equation}
These are the two orthogonal modes described in fig. \ref{fig:HOmodes}. 

We now define the reduced masses and frequencies of these modes:
\begin{eqnarray}
&m_r\equiv\frac{m_1}{2}& \qquad  \omega_r^2\equiv\frac{V_{ho}}{m_r}((\vec{\lambda_1}\cdot\vec{\lambda_2}) + \frac{1}{2}(\vec{\lambda_1}\cdot\vec{\lambda_3})) \nonumber \\
&m_R\equiv\frac{2m_1m_3}{2m_1+m_3}& \qquad \omega_R^2\equiv\frac{V_{ho}}{m_R}2(\vec{\lambda_1}\cdot\vec{\lambda_3})
\end{eqnarray}

The new Hamiltonian is a sum of two commuting Hamiltonians; therefore the wave
function will be a product of the eigenstates of these Hamiltonians:
\begin{equation}
\psi_0(\vec{R},\vec{r})=\psi_{R0}(\vec{r})\psi_{r0}(\vec{r})=(\frac{\alpha_R
\alpha_r}{\pi})^\frac{3}{2}e^{-\frac{1}{2}(\alpha_R^2R^2+\alpha_r^2r^2)}
\end{equation}
using the definition given above -- $\alpha_i\equiv \sqrt{\frac{m_i\omega_i}{\hbar}}$.

The calculation of the $\delta$-function expectation value is as follows:
\begin{eqnarray}
\langle \psi_0 | \delta(\vec{r_{13}}) | \psi_0 \rangle 
&=& \int \int d^3rd^3R\psi_0^*(\vec{R},\vec{r})\psi_0(\vec{R},\vec{r})\delta(\vec{R}-\frac{\vec{r}}{2}) \nonumber \\
&=& \int d^3r|\psi_{R0}(\frac{\vec{r}}{2})|^2|\psi_{r0}(\vec{r})|^2 \nonumber \\
&=& \Big(\frac{\alpha_r\alpha_R}{\pi}\Big)^3\Big(\int dx e^{-\alpha_R^2\frac{x^2}{4}-\alpha_r^2x^2}\Big)^3 \nonumber \\
&=& \Big(\frac{1}{\sqrt{\pi}}\frac{\alpha_r\alpha_R}{\sqrt{\frac{\alpha_R^2}{4}+\alpha_r^2}}\Big)^3
\end {eqnarray}

We can now plug the masses from (\ref{formula_quark_masses}) into the expressions we reached, and compute the HF splitting ratios for the meson and baryons under consideration (table \ref{tab_HO}). 

\begin{table} [!htbp]
\centering
\begin{tabular}{cccc} \hline
      Meson & Baryon    &  HF splitting &Deviation \\ 
      			&           & ratio         &from data      \\ \hline \hline 
\mystrut $K$&$\Sigma  $ & $1.65$        & $21\%$\\  
\mystrut $D$&$\Sigma_c$ & $1.62$        & $26\%$\\ 
\mystrut $B$&$\Sigma_b$ & $1.59$        & $26\%$\\ \hline 
\end{tabular}
\caption{\small{Prediction of meson / baryon HF splitting ratios for the harmonic oscillator model.}}
\label{tab_HO}
\end{table}

\subsection{Coulomb interaction}
The three-body problem with Coulomb interactions has no analytic solution, and we have to use variational methods in order to calculate the contact probabilities. The ansatz we chose was introduced by Coolidge and James
\cite{CoolidgeJames37,deSaavedra98} as part of a modification of the so-called Hylleraas method, and it uses the isoperimetrical coordinates:
\begin{eqnarray}
u &=& r_{12} + r_{23} - r_{13} \nonumber \\
v &=& r_{12} - r_{23} + r_{13} \nonumber \\
w &=& - r_{12} + r_{23} + r_{13}
\end{eqnarray}
The integration measure in these coordinates is:
\begin{eqnarray}
dr^3_{13}dr^3_{23} = \frac{\pi^2}{4} (u+v)(u+w)(w+v)du dv dw \nonumber 
\end{eqnarray}
and the kinetic terms take the form:
\begin{eqnarray}
\frac{p^2_1}{2m_1} = 
-\frac{2}{m_1(u+v)(v+w)}&\Bigg[& uw\Big(\frac{\partial^2}{\partial u^2}
									  	+\frac{\partial^2}{\partial w^2}-2\frac{\partial^2}{\partial u \partial w}\Big) \\ \nonumber
													  	&&	+~v(u+v+w)\frac{\partial^2}{\partial v^2} 
                             +(w-u)\Big(\frac{\partial}{\partial u}-\frac{\partial}{\partial w}\Big) \\ \nonumber
                                && +~(u+2v+w)\frac{\partial}{\partial v}\Bigg]
\end{eqnarray}
The kinetic terms for the second and third particles can be obtained by a permutation of the parameters $u$,$v$ and $w$.

On the basis of the solution for the 2 body problem with Coulomb interaction, and the requirement for a wave function which is symmetric under the exchange of the two light quarks (the $\Sigma$ has isospin-1), we choose the following ansatz:
\begin{eqnarray}
\psi(u,v,w)&&=e^{-\frac{1}{2}(\alpha(u+v)+\beta(w))}\times \\ \nonumber  
&&\sum_{k,l,m}C_{klm}\Big[L_k(\alpha u)L_l(\alpha v)+L_l(\alpha u)L_k(\alpha v)\Big]L_m(\beta w)
\label{formula_3body_ansatz}
\end{eqnarray}
where $\alpha$ , $\beta$ and $C_{klm}$ are variational parameters, and $L_k(x)$ are Laguerre polynomials of degree $[k,0]$. 

The extraction of the parameters was performed using the Ritz variational
method. Summing up to polynomials of rank $N\equiv k+l+m=5$, we were able to reproduce calculations 
of given in \cite{deSaavedra98} with a deviation of less than 1\%.

Given the ground state wave function we can calculate the contact probability
\begin{equation}
\langle \psi_0 | \delta(r_{us}) | \psi_0 \rangle_{baryon}=\frac{\pi}{2}\int v^2 dv |\psi (0,v,0)|^2
\end{equation}
leading to the results shown in table \ref{tab_Coulomb}.
\begin{table} [!htbp]
\centering
\begin{tabular}{cccc} \hline
      Meson & Baryon    &  HF splitting &Deviation \\ 
      			&           & ratio         &from data      \\ \hline \hline 
\mystrut $K$&$\Sigma  $ & $5.07\pm0.08$ & $144\%$\\  
\mystrut $D$&$\Sigma_c$ & $5.62\pm0.02$ & $158\%$\\ 
\mystrut $B$&$\Sigma_b$ & $5.75\pm0.01$ & $167\%$\\ \hline 
\end{tabular}
\caption{\small{Prediction of meson / baryon HF splitting ratios for the Coulomb interaction model.
The error bars are obtained by considering a 5\% error in the quark mass.}}
\label{tab_Coulomb}
\end{table}

\subsection{Linear potential}
The eigenstates of a 2-body Hamiltonian with a linear potential 
\begin{equation}
V_{linear}(r)={r}/{a^2}
\end{equation}
and reduced mass $\mu=1$ are 
\begin{equation}
\psi_n(r)=\frac{1}{r}Ai\Big[\frac{2r-2a^2E_n}{(2a)^{{2}/{3}}}\Big]
\end{equation}
where $Ai$ is the Airy function. The energy $E_n$ is determined by the boundary condition that the Airy function should equal zero as $r\rightarrow0$:
\begin{equation}
E_n=-\frac{a_n}{2^{1/3}a^{4/3}}
\end{equation}
with $a_n$ the $n$'th zero of the Airy function. 

\begin{figure}[!htbp]
\centering
\subfigure[] 
{
    \label{fig:airy:a}
    {\includegraphics[width=0.9\textwidth]{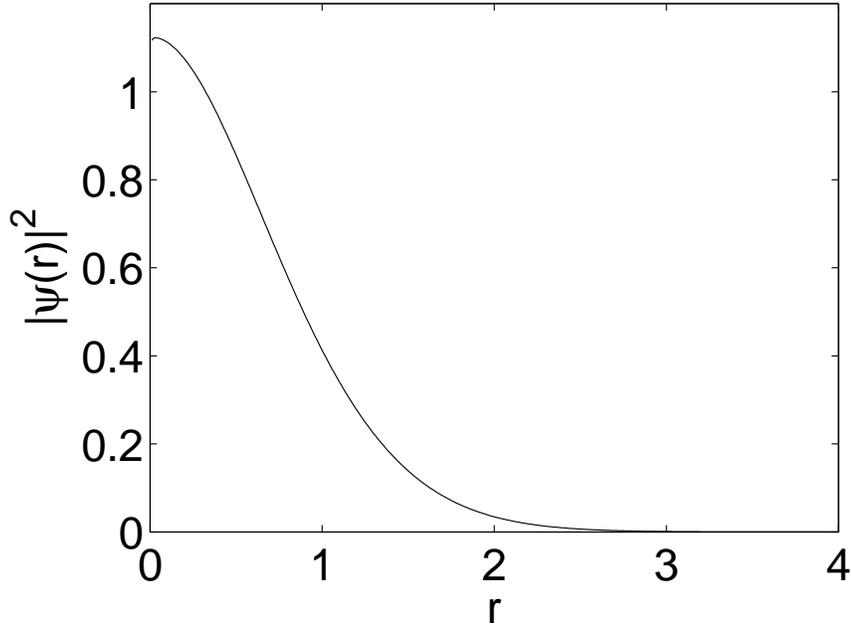}}
}
\subfigure[]
{
    \label{fig:airy:b}
    \includegraphics[width=0.9\textwidth]{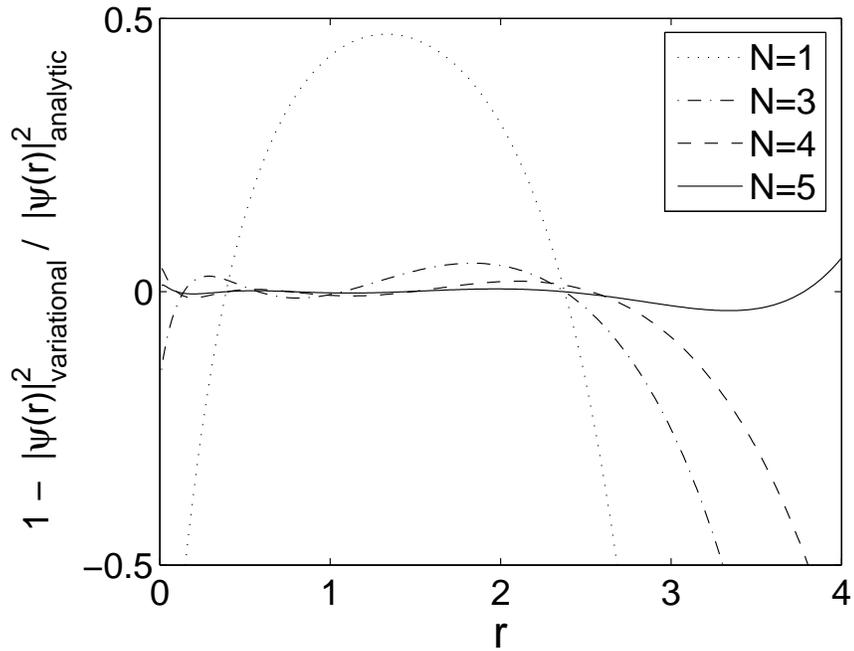}
}
\caption{\small{Plot (a) shows the analytic solution for a 2 body ground state
with a linear potential ($r$ is given in the units where $c=\hbar=\mu=a=1$). 
Plot (b) shows the accuracy of the solutions reached using the variational
method ($1-|\psi(r)_{variational}|^2/|\psi(r)_{analytic}|^2$) with different
polynomial degrees $N$.}}
\label{fig:sub} 
\end{figure}

Once again, the 3-body problem requires usage of variational methods. 
In order to avoid the difficulty of performing many integrations of Airy functions and their derivatives, we tried to use the modified Hylleraas ansatz (Eq. \ref{formula_3body_ansatz}) for the linear potential problem.
Testing this ansatz for the 2-body problem, with the wavefunction given by
\begin{equation}
\psi(r)=\sum_{k=1}^{N} C_ke^{-\frac{1}{2}\alpha r}L_k(\alpha r)~~,
\end{equation}
we found out that it was enough to take polynomials of degree $N_{max}=5$ for the variational solution to converge to the analytic one (see plot \ref{fig:airy:b}).

Following the success with the 2 body problem we felt confident enough to use the same software to calculate the contact probability ratio between mesons and baryons (see table \ref{tab_linear}).
\begin{table} [!htbp]
\centering
\begin{tabular}{cccc} \hline
      Meson & Baryon    &  HF splitting &Deviation \\ 
      			&           & ratio         &from data      \\ \hline \hline 
\mystrut $K$&$\Sigma  $ & $1.88\pm0.06$ & $14\%$\\  
\mystrut $D$&$\Sigma_c$ & $1.88\pm0.08$ & $10\%$\\ 
\mystrut $B$&$\Sigma_b$ & $1.86\pm0.09$ & $13\%$\\ \hline 
\end{tabular}
\caption{\small{Prediction of meson / baryon HF splitting ratios for the linear potential model.
The error estimates are based on comparison of the $N_{max}=5$ results with the analogous results for $N_{max}=4$.}}
\label{tab_linear}
\end{table}

\subsection{Cornell potential - Coulomb + linear}
\begin{equation}
V_{Cornell}(r)=-\frac{\kappa}{r}+\frac{r}{a^2}
\end{equation}
The Coulomb + linear potential combines the behavior of the color interaction in the two asymptotic limits, and it is theoretically supported by calculations using several techniques 
\cite{Brambilla94,sumino03}.
This model was used by the Cornell group \cite{Cornell78} and proved very successful in explaining various aspects of the hadronic spectrum.

Unlike the previous models, in this case the interaction strength does not cancel, and the HF splitting ratio depends both on the mass ratio $m_3/m_1$ and on an additional parameter:
\begin{equation}
K \equiv \kappa(m_ua)^{\frac{2}{3}}
\end{equation}
The parameters extracted from the charmonium spectrum in
\cite{Cornell80} are equivalent to the range $ 0.25 < K < 0.45$. 
$a$ and $\kappa$ include the product of the $SU(3)$ generators;
therefore their values are different for mesons and baryons.  Using Eq.\
(\ref{formula_SU3_generators}) we reach the relations
$ \kappa_{baryon}=\frac{1}{2}\kappa_{meson}$ and $a_{baryon}=\sqrt{2}a_{meson}$.

Once again, we tried to solve the 2 body problem using the Coulomb ansatz, with satisfactory results -- we were able to reproduce calculations for meson expectation values given in \cite[table I]{Cornell80} with deviation less than 1\% ($N_{max}=5$). Using this method to calculate the meson/baryon HF splitting ratio for the Cornell potential with $ 0.2 < K < 0.5$ we got the following results:
\begin{center}
	\centering
		\begin{tabular}{ccccc} 
\mystrut $2.04$ & $<$ & ${\Delta_K}/{\Delta_{\Sigma}}  $&$<$& $2.28$ \\ 
\mystrut $2.08$ & $<$ &${\Delta_D}/{\Delta_{\Sigma_c}} $&$<$& $2.40$ \\ 
\mystrut $2.07$ & $<$ &${\Delta_B}/{\Delta_{\Sigma_b}} $&$<$& $2.43$ \\ 
		\end{tabular}
\end{center}
The results for the specific value of $K=0.28$ are given in Table \ref{tab_Cornell}.

\begin{table} [!htbp]
\centering
\begin{tabular}{cccc} \hline
      Meson & Baryon    &  HF splitting &Deviation \\ 
      			&           & ratio         &from data      \\ \hline \hline 
\mystrut $K$&$\Sigma  $ & $2.10\pm0.05$ & $1\%$\\  
\mystrut $D$&$\Sigma_c$ & $2.16\pm0.07$ & $1\%$\\ 
\mystrut $B$&$\Sigma_b$ & $2.17\pm0.08$ & $1\%$\\ \hline 
\end{tabular}
\caption{\small{Prediction of meson / baryon HF splitting ratios for the Coulomb + linear potential model with $K=0.28$.
The error estimates are based on comparison of the $N_{max}=5$ results with the analogous results for $N_{max}=4$.}}
\label{tab_Cornell}
\end{table}

\subsection{Logarithmic potential}
\begin{equation}
V_{log}(r)=C\log(\frac{r}{r_0})
\end{equation}
One of the consequences of taking the logarithm as the function that describes the confining potential is that the energy level spacings are independent of the constituent quark masses. This phenomenon is observed when comparing 
the charmonium and the bottomonium spectra, and this is one of the strongest reasons why the logarithmic potential is considered a strong candidate for the confining potential.

In order to compute the HF splitting ratio with this potential, we used the same technique described above for the Coulomb and linear potentials. The theoretical results for the $1S$ quarkonium state given in \cite[tables 6-7]{QuiggRosner79} were accurately reproduced.

Unfortunately, when using the modified Hylleraas ansatz (\ref{formula_3body_ansatz}) the computation of the matrix elements $\langle \psi_i(\alpha,\beta) | \log(r) |\psi_j(\alpha,\beta)\rangle$ requires 
careful treatment at the limit $(\alpha-\beta)\rightarrow0$, otherwise the integrals diverge.
In order to avoid this we chose to take into account only a single variational parameter (i.e. we assumed $\alpha=\beta$).
In order to test whether this modification might generate large errors, we checked its effect on the results of the previous models, and the corrections did not exceed 1\%. The final results for the logarithmic potential 
are given in table \ref{tab_log}.

\begin{table} [!htbp]
\centering
\begin{tabular}{cccc} \hline
      Meson & Baryon    &  HF splitting &Deviation \\ 
      			&           & ratio         &from data      \\ \hline \hline 
\mystrut $K$&$\Sigma  $ & $2.38\pm0.02$ & $14\%$\\  
\mystrut $D$&$\Sigma_c$ & $2.43\pm0.02$ & $11\%$\\ 
\mystrut $B$&$\Sigma_b$ & $2.43\pm0.01$ & $13\%$\\ \hline 
\end{tabular}
\caption{\small{Prediction of meson / baryon HF splitting ratios for the logarithmic potential model.
The error estimates are based on comparison of the $N_{max}=5$ results with the analogous results for $N_{max}=4$.}}
\label{tab_log}
\end{table}

\section{Conclusions}
In this research we tried to explain the measured ratio between the HF splittings of mesons and baryons using the constituent quark model.
Thanks to the simple fact that the $\Sigma$ baryon is an isospin-1 particle with quark content similar to that of a $K$ meson, most of the degrees of freedom in the Sakharov-Zeldovich model are eliminated, 
and the HF splitting ratio was shown to depend only on the quarks' contact probabilities. 

Assuming different confining potentials we calculated the expected values for this measurement using variational methods. The software that implements these calculations reproduced similar calculations found in the literature with a deviation less than 1\%.

\begin{figure*}
\centering
{
{\includegraphics[width=0.72\textwidth]{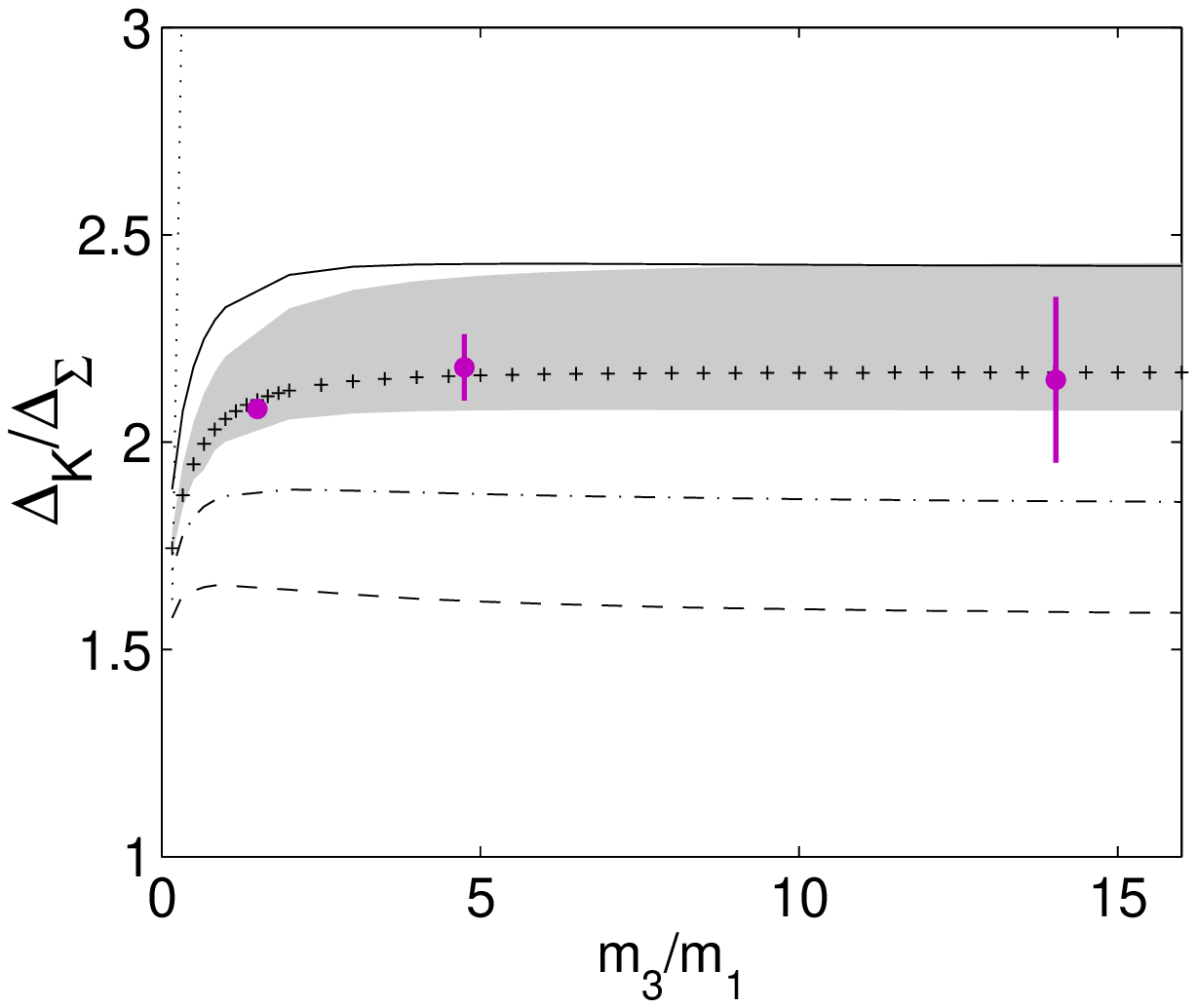}}
		 \includegraphics[width=0.22\textwidth,viewport=70 5 216 320,clip]{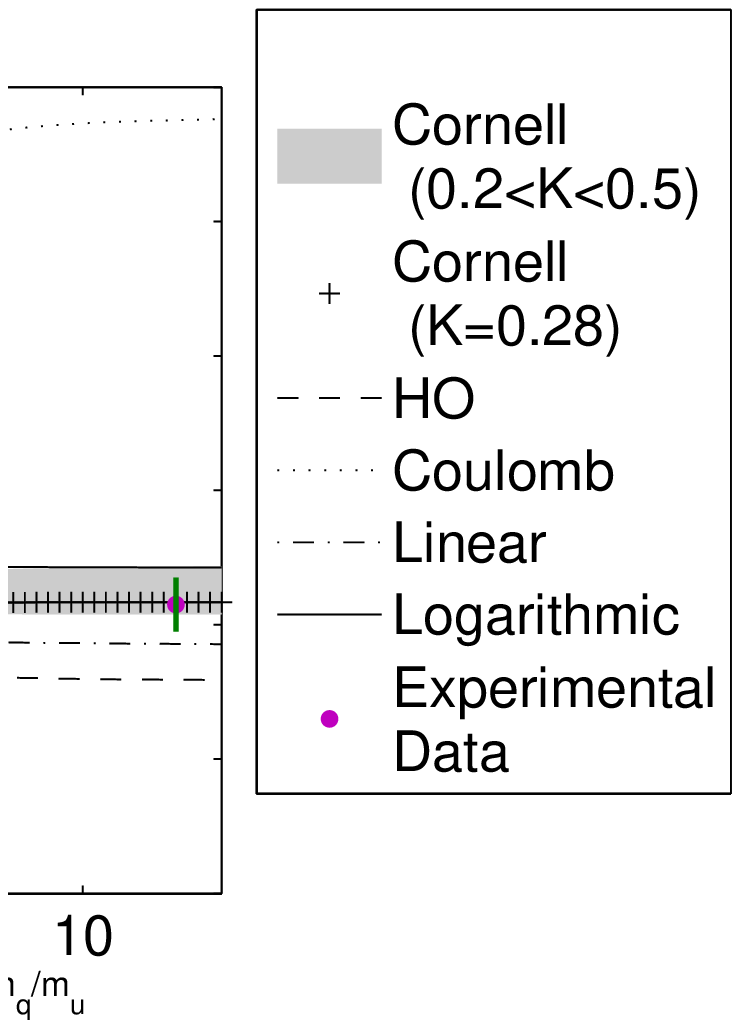}	
    \label{fig:conclusions:b}    
}
\caption{\small{The $K$-$\Sigma$ HF splitting ratio as a function of the quark
mass ratio assuming the different confining potentials. 
Most of the Coulomb model graph is outside the range of this plot.
The area marked in gray shows the results that correspond to a Cornell potential with $K$ values between 0.2 and 0.5. 
One can see the good fit between the data and the predictions obtained 
with the Cornell potential (K=0.28).}}
\label{fig:conclusions} 
\end{figure*}

The results, which are summarized in Fig. \ref{fig:conclusions}, show that this
quantity enables us to distinguish between the different confinement models,
and that the experimental data favor a Cornell potential, similarly to the
results obtained by previous experiments.

\section*{Acknowledgements}
I would like to thank Marek Karliner, Harry Lipkin and Jon Rosner for 
encouragement and valuable discussions. This research was supported by a grant from the
Israel Science Foundation administered by the Israel Academy of Sciences and Humanities.

\end{document}